\documentstyle[12pt]{article}

\begin{document}

\noindent {\small USC-99/HEP-B1\hfill \hfill hep-th/9904063}\newline
{\small \hfill }

{\vskip 2cm}

\begin{center}
{\Large {\bf Lifting M-theory to Two-Time Physics}}{\footnote{
This research was partially supported by the US. Department of Energy under
grant number DE-FG03-84ER40168.}}{\Large {\bf \ }}

\bigskip

{\vskip 1cm}

{\bf Itzhak Bars, Cemsinan Deliduman, Djordje Minic}

{\vskip 1cm}

{Department of Physics and Astronomy}

{University of Southern California}

{\ Los Angeles, CA 90089-0484, USA}

{\vskip 2cm}

{\bf Abstract}

{\vskip 1cm}
\end{center}

M-theory has different global supersymmetry structures in its various dual
incarnations, as characterized by the M-algebra in 11D, the type IIA,
type-IIB, heterotic, type-I extended supersymmetries in 10D, and non-Abelian
supersymmetries in the AdS$_n\times S^m$ backgrounds. We show that all of
these supersymmetries are unified within the supersymmetry OSp$\left(
1/64\right) $, thus hinting that the overall global spacetime symmetry of
M-theory is OSp$\left( 1/64\right) $. We suggest that the larger symmetries
contained within OSp$\left( 1/64\right) $ which go beyond the familiar
symmetries, are non-linearly realized hidden symmetries of M-theory. These
can be made manifest by lifting 11D M-theory to the formalism of two-time
physics in 13D by adding gauge degrees of freedom. We illustrate this idea
by constructing a toy M-model on the worldline in 13D with manifest OSp$%
\left( 1/64\right) $ global supersymmetry, and a number of new local
symmetries that remove ghosts. Some of the local symmetries are bosonic
cousins of kappa supersymmetries. The model contains 0-superbrane and
p-forms (for p=3,6) as degrees of freedom. The gauge symmetries can be fixed
in various ways to come down to a one time physics model in 11D, 10D, AdS$%
_n\times S^m$, etc., such that the linearly realized part of OSp$\left(
1/64\right) $ is the global symmetry of the various dual sectors of M-theory.

\newpage

\section{Perspective on hidden symmetries}

M-theory is defined in 11 dimensions, with one time and ten space
coordinates. It has an extended global supersymmetry characterized by 32
supercharges $Q_\alpha $ and 528 {\it abelian} bosonic charges that include
the momentum $P_\mu $, the two-brane charge $Z_{\mu \nu }$ and the
five-brane charge $Z_{\mu _1\cdots \mu _5}$ \cite{townsend}-\cite{gunaydin} 
\begin{equation}
\left\{ Q_\alpha ,Q_\beta \right\} =\gamma _{\alpha \beta }^\mu P_\mu
+\gamma _{\alpha \beta }^{\mu \nu }Z_{\mu \nu }+\gamma _{\alpha \beta }^{\mu
_1\cdots \mu _5}Z_{\mu _1\cdots \mu _5}\,.  \label{M-algebra}
\end{equation}
The charges $P_\mu $, $Z_{\mu \nu }$, $Z_{\mu _1\cdots \mu _5}$ commute
among themselves and with $Q_\alpha $. In addition, the 11-dimensional SO$%
\left( 10,1\right) $ Lorentz generator $J_{\mu \nu }$ has non-trivial
commutation rules with $Q_\alpha ,$ $P_\mu ,$ $Z_{\mu \nu },$ $Z_{\mu
_1\cdots \mu _5}$ that correspond to the classification of these charges as
spinor, 1-form, 2-form, 5-form respectively in 11-dimensions. We refer to
the algebra satisfied by $J_{\mu \nu }$, $Q_\alpha $, $P_\mu $, $Z_{\mu \nu
} $, $Z_{\mu _1\cdots \mu _5}$ as the M-algebra.

It is well known that physical input in four dimensions, such as the absence
of massless interacting particles with spin higher than two, constrains the
maximum number of supersymmetries to 32 \cite{nahm}. This refers to the
maximum number of $Q_\alpha $ that commute with the momentum operators $%
P_\mu $, as is the case in the M-algebra. However, there is no physical
restriction on the number of non-linearly realized supersymmetries that do
not commute with $P^\mu $. In particular, it is possible to have
supersymmetries that do not commute with the Hamiltonian or $P^\mu $, while
being supersymmetries of the action. For example, this is the case with the
well known special superconformal symmetry generated by the fermions $%
S_\alpha $ in any superconformal theory in any dimension. In addition, it
has been discovered recently that very simple familiar systems have
previously unnoticed hidden symmetries of the type SO$\left( d,2\right) $,
that are not symmetries of the Hamiltonian \cite{lifting} \cite{super2t} but
are symmetries of the action. Such symmetries are made manifest by lifting
the system to the formalism of two-time physics by the addition of gauge
degrees of freedom together with new gauge symmetries \cite{lifting} \cite
{super2t}.

In this paper we provide arguments that the supersymmetries of the different
dual versions of M-theory are all unified within OSp$\left( 1/64\right) $,
with 64 supercharges. We will show that the M-algebra is a subalgebra of OSp$%
\left( 1/64\right) $ without any contractions. The extra 32 supercharges do
not commute with $P^\mu $ within OS$p\left( 1/64\right) $. We suggest they
are symmetries of the ``action'' of M-theory. The symmetry structures that
emerge in this way suggest that M-theory could admit a two-time physics
formulation with a total of 13 dimensions.

There has been a number of hints that M-theory may contain various two-time
structures \cite{duff},\cite{ibjapan},\cite{vafa}-\cite{nishino}. In this
paper we will show a new embedding of the symmetries of M-theory in a higher
structure suggested by the formalism of two-time physics. We show that when
the M-algebra is extended by adding the 11D conformal generator $K^\mu $,
the closure requires the full OSp$\left( 1/64\right) $. Duality and 11D
covariance suggest that $K^\mu \,$is a hidden symmetry in M-theory. The
point is that the 10D covariant type-IIB superalgebra, as well as heterotic
and type I superalgebras can be obtained from the same OSp$\left(
1/64\right) $ , and the 10D type-IIA superalgebra is just the dimensional
reduction of the M-theory algebra. While $K^\mu \,$is non-linearly realized
and remains ``hidden'' in the 11D version of M-theory, some of its
components are linearly realized in a dual version of M-theory, so $K^\mu $
is actually present non-perturbatively. A further clue is the presence of
conformal symmetry in some corners of M-theory as noted through the CFT-AdS
correspondance \cite{maldacena}-\cite{gunayd}.

These points will be illustrated in a specific toy M-model \cite{future}
with a worldline action that includes 13D p-form degrees of freedom $a^{%
\tilde{M}_1\cdots \tilde{M}_p}\left( \tau \right) $ for $p=3,6$ in addition
to the zero brane degrees of freedom $X^{\tilde{M}}\left( \tau \right) $, $%
P^{\tilde{M}}\left( \tau \right) $, $\Theta ^{\tilde{\alpha}}\left( \tau
\right) $ that normally exist in a worldline formalism \cite{super2t}. The
model introduces new concepts of local symmetries, including one that is a
bosonic cousin of kappa supersymmetry. Various gauge choices can be found to
yield the M-algebra in 11D, the type IIA, type-IIB, heterotic, type-I
extended supersymmetries in 10D, and non-Abelian superalgebras in the AdS$%
_n\times S^m$ backgrounds. Thus, the symmetries of different corners of the
moduli space of M-theory emerge as different gauge choices in this model 
\cite{future}.

\section{Subgroup chains}

We first show that the M-algebra is contained in OSp$\left( 1/64\right) $.
The supergroup OSp$\left( 1/64\right) $ has 64 fermionic generators $Q_{%
\tilde{\alpha}}$ and $\allowbreak 2080$ bosonic generators $S_{\tilde{\alpha}%
\tilde{\beta}}$ that form a 64$\times $64 symmetric matrix. The Lie
superalgebra is 
\begin{eqnarray*}
\left\{ Q_{\tilde{\alpha}},Q_{\tilde{\beta}}\right\}  &=&S_{\tilde{\alpha}%
\tilde{\beta}},\quad \left[ S_{\tilde{\alpha}\tilde{\beta}},Q_{\tilde{\gamma}%
}\right] =C_{\tilde{\alpha}\tilde{\gamma}}Q_{\tilde{\beta}}+C_{\tilde{\beta}%
\tilde{\gamma}}Q_{\tilde{\alpha}}\, \\
\left[ S_{\tilde{\alpha}\tilde{\beta}},S_{\tilde{\gamma}\tilde{\delta}%
}\right]  &=&C_{\tilde{\alpha}\tilde{\gamma}}S_{\tilde{\beta}\tilde{\delta}%
}+C_{\tilde{\alpha}\tilde{\delta}}S_{\tilde{\beta}\tilde{\gamma}}+C_{\tilde{%
\beta}\tilde{\gamma}}S_{\tilde{\alpha}\tilde{\delta}}+C_{\tilde{\beta}\tilde{%
\delta}}S_{\tilde{\alpha}\tilde{\gamma}}\,
\end{eqnarray*}
The $S_{\tilde{\alpha}\tilde{\beta}}$ form the Lie algebra of Sp$\left(
64\right) $ and the constant antisymmetric matrix $C_{\tilde{\alpha}\tilde{%
\beta}}$ is the metric of Sp$\left( 64\right) $. A matrix representation of
OSp$\left( 1/64\right) $ is given by 65$\times $65 supermatrices. We are
interested in re-expressing these generators in various bases, that are
related to the basis above by unitary transformations, such that various
spacetime interpretations can be given to those bases. For this purpose
several branchings of Sp$\left( 64\right) $ will be used : 
\begin{eqnarray}
A &:&Sp\left( 64\right) \supset SU^{*}\left( 32\right) \otimes SO\left(
1,1\right)   \label{A} \\
&\supset &Sp^{*}\left( 32\right) \otimes SO\left( 1,1\right) \supset
SO\left( 10,2\right) \otimes SO\left( 1,1\right) \,\supset \cdots   \nonumber
\\
B &:&Sp\left( 64\right) \supset SO^{*}\left( 32\right) \otimes SO\left(
2,1\right)   \label{B} \\
&\supset &SU^{*}\left( 16\right) \otimes U\left( 1\right) \otimes SO\left(
2,1\right)   \nonumber \\
&\supset &SO\left( 9,1\right) \otimes SO\left( 2,1\right) \otimes U\left(
1\right) \,\supset \cdots   \nonumber \\
C &:&Sp\left( 64\right) \supset SO\left( 11,2\right) \supset \cdots 
\label{C}
\end{eqnarray}
The (*) indicates an appropriate analytic continuation that contains the
non-compact groups listed. There is another chain of interest that involves
a supergroup that will come up in our discussion (we are not listing a
complete set of branchings). 
\begin{equation}
OSp\left( 1/64\right) \supset OSp\left( 1/32\right) \otimes Sp\left(
32\right) \supset \cdots 
\end{equation}
We list the first step of the decomposition of the 64$\oplus $2080
representations for the A,B,C branches. For the A-branch we have the $%
SU^{*}\left( 32\right) \otimes SO\left( 1,1\right) $ representations  
\begin{eqnarray}
64 &=&32^{1/2}\oplus \overline{32}^{-1/2},  \label{A1} \\
2080 &=&\left( 1^0\oplus 1023^0\,\right) _{32\cdot \overline{32}}\,\oplus
528_{\left( 32\cdot 32\right) _s}^{+}\oplus 528_{\left( \overline{32}\cdot 
\overline{32}\right) _s}^{-}  \label{A2}
\end{eqnarray}
where the superscripts correspond to the SO$\left( 1,1\right) $ charge, and
subscripts indicate the products of the supercharges that produce those
representations ($s$ and $a$ stand for symmetric and antisymmetric product
respectively). For the B-branch we have the $SO^{*}\left( 32\right) \otimes
SO\left( 2,1\right) $ representations  
\begin{eqnarray}
64 &=&\left( 32,2\right) ,  \label{B1} \\
2080 &=&\left( 1_{(32\cdot 32)_s},3_{\left( 2\cdot 2\right) _s}\right)
\,\,\,\,\oplus \left( 527_{(32\cdot 32)_s},3_{\left( 2\cdot 2\right)
_s}\right) \oplus \left( 496_{\left( 32\cdot 32\right) _a},1_{\left( 2\cdot
2\right) _a}\right) .  \label{B2}
\end{eqnarray}
For the C-branch we have the $SO\left( 11,2\right) $ representations  
\begin{equation}
64=64,\quad 2080=78\,\,(\tilde{J}_2)\oplus 286\,\,(\tilde{J}_3)\oplus
1716\,\,(\tilde{J}_6)  \label{C1}
\end{equation}
where the $\tilde{J}_{\tilde{M}\tilde{N}}$ , $\tilde{J}_{\tilde{M}_1\tilde{M}%
_2\tilde{M}_3}$, $\tilde{J}_{\tilde{M}_1\cdots \tilde{M}_6}$ are p-forms in
13D. The $\tilde{J}_p$ can be represented by p-products of 13D 64$\times 64$
gamma matrices. For $p=2,3,6$ these are 64$\times $64 symmetric matrices
that represent the 2080 components of $S_{\tilde{\alpha}\tilde{\beta}}$ in a
13D spinor basis. The A,B,C branches in this paper are related to the A,B,C
branches discussed in S-theory \cite{stheory}. The C branch provides a SO$%
\left( 11,2\right) $ covariant 13D interpretation for M-theory as we will
see with an explicit toy M-model.

\section{11D and 12D interpretations}

If SO$\left( 11,2\right) $ of the C-branch is interpreted as the conformal
group in 11-dimensions then one may re-classify all the generators as
representations of the Lorentz subgroups SO$\left( 10,1\right) \times
SO\left( 1,1\right) $ in 11-dimensions and 2-dimensions contained in 13D. To
do so, re-label the 13 dimensions with two sets, one in 10+1 and the other
in 1+1 dimensions, $\tilde{M}=\mu \oplus m$ where $\mu =0,1,\cdots ,10$ and $%
m=\left( +^{\prime },-^{\prime }\right) $, with the metric in the extra two
dimensions taken in a lightcone type basis $\eta ^{+^{\prime }-^{\prime
}}=-1 $. This conformal basis emerges naturally as one of the gauge choices
in two-time physics \cite{lifting} \cite{super2t} \cite{future} as will be
discussed later in this paper. The 64-spinor may be re-labelled as $Q_{%
\tilde{\alpha}}\sim Q_\alpha ^{1/2}\oplus S_\alpha ^{-1/2}$ with $\alpha $
denoting the 32-spinor in 11D and $\pm \frac 12$ denoting the two chiral
spinors in 2D. Then the generators $J^{\tilde{M}\tilde{N}}$ of the conformal
group are identified as $J^{\mu \nu }$-SO$\left( 10,1\right) $ Lorentz
transformations, $J^{+^{\prime }\mu }\equiv P^\mu $ -translations, $%
J^{-^{\prime }\mu }\equiv K^\mu $ -special conformal transformations, and $%
J^{+^{\prime }-^{\prime }}\equiv D$ -dilatations. All the generators have
definite dimensions under the commutation relations with $D$ which generates
the SO$\left( 1,1\right) $ subgroup. 
\begin{eqnarray}
D &=&1:\quad J^{+^{\prime }\mu }\equiv P^\mu ,\quad J^{+^{\prime }\mu \nu
}\equiv Z^{\mu \nu },\quad J^{+^{\prime }\mu _1\cdots \mu _5}\equiv Z^{\mu
_1\cdots \mu _5}  \label{D1} \\
D &=&\frac 12:\quad Q_{\alpha \frac 12}\equiv Q_\alpha  \label{D12} \\
D &=&0:\quad \left\{ 
\begin{array}{c}
J^{\mu \nu },\quad J^{+^{\prime }-^{\prime }}\equiv D,\quad \,J^{+^{\prime
}-^{\prime }\mu }\equiv J^\mu ,\,\,\quad J^{\mu _1\mu _2\mu _3},\,\, \\ 
J^{+^{\prime }-^{\prime }\mu _1\cdots \mu _4}\equiv J^{\mu _1\cdots \mu
_4},\,\,\quad J^{\mu _1\cdots \mu _6}\equiv \varepsilon ^{\mu _1\cdots \mu
_6\nu _1\cdots \nu _5}J_{\nu _1\cdots \nu _5}
\end{array}
\right\}  \label{D0} \\
D &=&\frac{-1}2:\quad Q_{\alpha \frac{-1}2}\equiv S_\alpha  \label{D-12} \\
D &=&-1:\quad J^{-^{\prime }\mu }\equiv K^\mu ,\quad J^{-^{\prime }\mu \nu
}\equiv \tilde{Z}^{\mu \nu },\quad J^{-^{\prime }\mu _1\cdots \mu _5}\equiv 
\tilde{Z}^{\mu _1\cdots \mu _5}  \label{D-1}
\end{eqnarray}
The dimensions $D=\pm 1,\pm \frac 12,0$ provide a 5-grading of the
superalgebra such that under the commutation rules the dimensions add $%
[X_{D_1},X_{D_2}\}\sim X_{D_1+D_2}$. If $D_1+D_2$ is not one of the
dimensions listed, the result of the commutator is zero. Then we see that
the $D\geq 0$ operators $J^{\mu \nu }$, $Q_\alpha ^{1/2}$, $P^\mu $, $Z^{\mu
\nu }$, $Z^{\mu _1\cdots \mu _5}$, form the M-algebra: the anti-commutator $%
\left\{ Q_\alpha ,Q_\beta \right\} $ contains only the 528 generators $P^\mu 
$, $Z^{\mu \nu }$, $Z^{\mu _1\cdots \mu _5}$ (D=1) which commute among
themselves and with $Q_\alpha ^{1/2}$ (D=1/2), while $J^{\mu \nu }$ (D=0)
generates the Lorentz transformations SO$\left( 10,1\right) $. Hence, the
11D M-algebra is contained in OSp$\left( 1/64\right) $ as a subalgebra
without considering any contractions!

OSp$\left( 1/64\right) $ is the smallest simple supergroup that includes the
M-algebra as a sub-algebra without resorting to contractions. As argued
above, if the special conformal generator $K^\mu $ is also included along
with the M-algebra, then all other generators of OSp$\left( 1/64\right) $ 
{\it must} also be included for consistency with the 5-grading and Jacobi
identities. Thus, conformal symmetry in 11D together with the M-algebra
demand OSp$\left( 1/64\right) $.

The 11D interpretation fits into the higher algebraic structures contained
in the A-branch (\ref{A}). The spinor decomposes as 64=32$^{1/2}\oplus 
\overline{32}^{-1/2}$ under SU$^{*}\left( 32\right) \otimes SO\left(
1,1\right) $. Each 32-spinor corresponds to the fundamental representation
of Sp$^{*}\left( 32\right) \subset $SU$^{*}\left( 32\right) $ and
furthermore they are classified as real Weyl spinors of SO$\left(
10,2\right) $ of {\it same} chirality. This classification defines the
A-envelop of the 5-graded 11D superconformal algebra described above such
that 11D is embedded in 12D. One may then combine the 11D operators $J_\mu
\oplus J_{\mu \nu }\,$ into a 12D generator $J_{MN}$, and the 5-form $J_{\mu
_1\cdots \mu _5}$ may be written as a 6-form $J_{M_1\cdots M_6}$ that is
self-dual in 12D. The $J_{MN}$ form the SO$\left( 10,2\right) $ subgroup
listed in the A-branch, and together with the $J_{M_1\cdots M_6}$ they make
up the 528-adjoint of Sp$\left( 32\right) $. Similarly, the remaining
operators make up complete 12D representations, such as $P_1\oplus
Z_2=Z_{MN} $ and $Z_5=Z_{M_1\cdots M_6}^{+}$ and $J_3\oplus J_4=J_{M_1\cdots
M_4}$, etc. Finally all of these are put together as SU$^{*}\left( 32\right)
\otimes SO\left( 1,1\right) $ representations as in the first step of the
A-branch in (\ref{A1}-\ref{A2}). This shows that the operators $J_{MN}$, $%
Z_{MN},$ $Z_{M_1\cdots M_6}^{+},Q_\alpha $ which form a 12D envelop for the
M-algebra, as used in several applications of S-theory, also fit in the
two-time formalism given in this paper.

\section{Type-IIA, IIB, heterotic, type-I interpretations}

The 10D type-IIA version follows from re-classifying the 11D basis of (\ref
{D1}-\ref{D-1}) under SO$\left( 9,1\right) $ of 10D. The 32-spinor
supercharge $Q_\alpha ^{1/2}$ in 11D becomes the two opposite chirality
supercharges 16+$\overline{16}$ of type IIA. Similarly the 32-spinor of
special superconformal generator $S_\alpha ^{-1/2}$ becomes the two opposite
chirality special superconformal supercharges 16+$\overline{16}$ of type
IIA. The 11D M-algebra is rewritten trivially in 10D in the type IIA basis.

In the C/B-branch, the type-IIB version follows from re-classifying the
64-spinor of 13D as the spinor$\times $spinor of 10D$\oplus $3D, namely $Q_{%
\tilde{\alpha}}=Q_{\alpha a}\oplus S_{\dot{\alpha}a}$ where $\alpha ,\dot{%
\alpha}$ denote the real spinors 16,$\overline{16}$ of SO$\left( 9,1\right) $
and $a$ denotes the doublet spinor of SO$\left( 2,1\right) $. Both sets of
spinors $Q_{\alpha a}$, $S_{\dot{\alpha}a}$ are real, and there is no
relation between them via hermitian conjugation. The $Q_{\alpha a}$ play the
role of the two 10D supersymmetry generators of type IIB, while the two $S_{%
\dot{\alpha}a}$ play the role of the two 10D special superconformal
generators of type IIB. Similarly the vector of SO$\left( 11,2\right) $ is
decomposed by using $\tilde{M}=\bar{\mu}+\bar{m}$ with $\bar{\mu}=0,1,\cdots
,9$ and $\bar{m}=+^{\prime },-^{\prime },2^{\prime }$. The $\bar{m}=\pm
^{\prime }$ components are the same as the $\pm ^{\prime }$ components used
to reduce 13D to 11D, and the $2^{\prime }$ component is a re-naming of the
11th dimension in SO$\left( 10,1\right) $. The anti-commutators in the
superalgebra take the sketchy form 
\begin{eqnarray}
\left\{ Q_{\alpha a},Q_{\beta b}\right\} &=&P_{\bar{\mu}}^{\bar{m}}\oplus Z_{%
\bar{\mu}_1\cdots \bar{\mu}_5}^{\bar{m}}\oplus Z_{\bar{\mu}_1\bar{\mu}_2\bar{%
\mu}_3} \\
\left\{ S_{\dot{\alpha}a},S_{\dot{\beta}b}\right\} &=&\,K_{\bar{\mu}}^{\bar{m%
}}\oplus \tilde{Z}_{\bar{\mu}_1\cdots \bar{\mu}_5}^{\bar{m}}\oplus \tilde{Z}%
_{\bar{\mu}_1\bar{\mu}_2\bar{\mu}_3} \\
\left\{ Q_{\alpha a},S_{\dot{\beta}b}\right\} &=&D_{\bar{m}\bar{n}}\oplus J_{%
\bar{\mu}_1\bar{\mu}_2}^{\bar{m}}\oplus X_{\bar{\mu}_1\cdots \bar{\mu}_4}^{%
\bar{m}}\oplus \chi \oplus J_{\bar{\mu}_1\bar{\mu}_2}\oplus X_{\bar{\mu}%
_1\cdots \bar{\mu}_4}
\end{eqnarray}
The map between the 13D notation and the 10D+3D notation follows from $%
\tilde{M}=\bar{\mu}\oplus \bar{m}$ 
\begin{eqnarray}
Q_{\tilde{\alpha}} &\sim &Q_{\alpha a}\oplus S_{\dot{\alpha}a} \\
J_{\tilde{M}_1\tilde{M}_2} &\sim &J_{\bar{\mu}_1\bar{\mu}_2}\oplus \left( P_{%
\bar{\mu}}^{\bar{m}}+K_{\bar{\mu}}^{\bar{m}}\right) \oplus D_{\bar{m}_1\bar{m%
}_2} \\
J_{\tilde{M}_1\tilde{M}_2\tilde{M}_3} &\sim &\left( Z_{\bar{\mu}_1\bar{\mu}_2%
\bar{\mu}_3}-\tilde{Z}_{\bar{\mu}_1\bar{\mu}_2\bar{\mu}_3}\right) \oplus J_{%
\bar{\mu}_1\bar{\mu}_2}^{\bar{m}}\oplus \left( P_{\bar{\mu}}^{\bar{n}}-K_{%
\bar{\mu}}^{\bar{n}}\right) \varepsilon _{\bar{m}_1\bar{m}_2\bar{n}}\oplus
\chi \,\,\varepsilon _{\bar{m}_1\bar{m}_2\bar{m}_3} \\
J_{\tilde{M}_1\cdots \tilde{M}_6} &\sim &\varepsilon _{\bar{\mu}_1\cdots 
\bar{\mu}_6\bar{\nu}_1\cdots \bar{\nu}_4}X^{\bar{\nu}_1\cdots \bar{\nu}%
_4}\oplus \left[ Z_{\bar{\mu}_1\cdots \bar{\mu}_5}^{\bar{m}}\oplus Z_{\bar{%
\mu}_1\cdots \bar{\mu}_5}^{\bar{m}}\right] \\
&&\oplus X_{\bar{\mu}_1\cdots \bar{\mu}_4}^{\bar{n}}\varepsilon _{\bar{m}_1%
\bar{m}_2\bar{n}}\oplus \left( Z_{\bar{\mu}_1\bar{\mu}_2\bar{\mu}_3}+\tilde{Z%
}_{\bar{\mu}_1\bar{\mu}_2\bar{\mu}_3}\right) \varepsilon _{\bar{m}_1\bar{m}_2%
\bar{m}_3}
\end{eqnarray}
where $\oplus $ is direct sum, but $\pm $ imply ordinary addition or
subtraction, $\varepsilon _{\bar{m}_1\bar{m}_2\bar{m}_3}$ is the SO$\left(
2,1\right) $ invariant Levi-Civita tensor, and $Z_{\bar{\mu}_1\cdots \bar{\mu%
}_5}^{\bar{m}}$, $\tilde{Z}_{\bar{\mu}_1\cdots \bar{\mu}_5}^{\bar{m}}$ are
self-dual and anti self-dual respectively in 10D.

The SO$\left( 9,1\right) $ generators are $J_{\bar{\mu}_1\bar{\mu}_2}$ and
the SO$\left( 2,1\right) $ generators are $D_{\bar{m}\bar{n}}$. The
operators labelled with $\bar{m}$ are triplets of SO$\left( 2,1\right) $
while the others are singlets. The singlet operator $\chi $ is written in
terms of the 13D operators as $\chi =J^{+^{\prime }-^{\prime }2^{\prime }}$.
Its commutation rules with the 64 spinors $Q_{\tilde{\alpha}}$ is $\left[
\chi ,Q_{\tilde{\alpha}}\right] \sim \left( \Gamma ^{+^{\prime }-^{\prime
}2^{\prime }}Q\right) _{\tilde{\alpha}}$. Since one may write $\Gamma
^{+^{\prime }-^{\prime }2^{\prime }}$= $\Gamma ^{01\cdots 9}$ for 64$\times $%
64 gamma matrices \cite{stheory}, the operator $\chi $ acts like the
chirality operator on the 10D spinors $Q_{\alpha a}$, $S_{\dot{\alpha}a}$.
Therefore $\chi $ provides a 5-grading for the OSp$\left( 1/64\right) \,$%
operators based on their 10D chirality 
\begin{eqnarray}
\chi &=&1:\quad P_1^1\oplus Z_5^1\oplus Z_3  \label{c1} \\
\chi &=&\frac 12:\quad Q_{\alpha a}  \label{c12} \\
\chi &=&0:\quad \chi \oplus D^2\oplus J_2\oplus J_2^1\oplus X_4\oplus X_4^1
\label{c0} \\
\chi &=&\frac{-1}2:S_{\dot{\alpha}a}\quad  \label{c-12} \\
\chi &=&-1:\quad K_1^1\oplus \tilde{Z}_5^1\oplus \tilde{Z}_3  \label{c-1}
\end{eqnarray}
The bosonic generators $P_1^1,Z_5^1$ etc. are labelled by numbers in the
subscripts and superscripts $Z_p^q$ that correspond to $p$-forms in 10D and $%
q$-forms in 3D respectively. So the commutation rules for OSp$\left(
1/64\right) $ may be written in a graded chirality basis in the form $%
[X_{\chi _1},X_{\chi _2}\}=X_{\chi _1+\chi _2}$. If the chirality $\chi
_1+\chi _2$ does not exist the result of the commutator is zero. The $\chi
\geq 0$ operators $J_2$, $D^2$, $\chi ,$ $Q_{\alpha a},$ $P_1^1,$ $Z_5^1,$ $%
Z_3$ define the B-algebra. The chirality grading shows that $P_1^1,Z_5^1,Z_3$
commute with each other as well as with $Q_{\alpha a}$, while the SO$\left(
9,1\right) \otimes $SO$\left( 2,1\right) \otimes $U$\left( 1\right) $
subgroup, consisting of $J_2$, $D^2$, $\chi $, map the operators $Q_{\alpha
a},$ $P_1^1,$ $Z_5^1,$ $Z_3$ into themselves.

We have established that the B-algebra, that is essential for understanding
M-theory in a IIB basis, is included in OSp$\left( 1/64\right) $ as a
subalgebra without any contractions. Furthermore, since it fits into the
C-branch it is consistent with the SO$\left( 11,2\right) $ symmetry of
two-time physics and the idea that the 10D+3D basis is arrived at as a gauge
choice in two-time physics. Indeed this is true in the toy M-model discussed
later.

These C-branch 13D$=$10D+3D results intersect the B-branch 10D+3D basis as
can be seen by the following larger classification under the B-branch (\ref
{B}) that include the envelops SU$^{*}\left( 16\right) \times U\left(
1\right) $ and SO$^{*}\left( 32\right) \times U\left( 1\right) $%
\begin{eqnarray}
\left( 496,1\right) &=&\left[ \chi \oplus \left( J_2\oplus X_4\right)
\right] _{\left( 16\times \overline{16}\right) \left( 2\times 2\right)
_a}\oplus \left( Z_3\right) _{\left( 16\times 16\right) _a\left( 2\times
2\right) _a}\oplus \left( \tilde{Z}_3\right) _{\left( \overline{16}\times 
\overline{16}\right) _a\left( 2\times 2\right) _a} \\
\left( 1,3\right) &=&\left( D^2\right) _{\left( 16\times \overline{16}%
\right) _s\left( 2\times 2\right) _s} \\
\left( 527,3\right) &=&\left( P_1^1\oplus Z_2^1\oplus Z_5^1\right) _{\left(
16\times 16\right) s\left( 2\times 2\right) _s}\oplus \left( K_1^1\oplus 
\tilde{Z}_2^1\oplus \tilde{Z}_5^1\right) _{\left( \overline{16}\times 
\overline{16}\right) _s\left( 2\times 2\right) _s} \\
&&\oplus \left( J_2^1\oplus X_4^1\right) _{\left( 16\times \overline{16}%
\right) \left( 2\times 2\right) _s}  \nonumber
\end{eqnarray}
The products of SO$\left( 9,1\right) \otimes $ SO$\left( 2,1\right) $ spinor
representations $Q=\left( 16,2\right) $ and $S=\left( \overline{16},2\right) 
$ that produce the various 10D and 3D forms are indicted. The SU$^{*}\left(
16\right) \otimes U\left( 1\right) $ subgroup of SO$^{*}\left( 32\right) $
is generated by $\left( J_2\oplus X_4\right) \oplus \chi $. The forms in
independent parentheses $\left( \cdots \right) $ correspond to irreducible
representations under SU$^{*}\left( 16\right) \otimes SO\left( 2,1\right)
\otimes U\left( 1\right) $. Their collection in each line correspond to SO$%
^{*}\left( 32\right) \otimes SO\left( 2,1\right) $ representations as given
in (\ref{B2}). The $496$ is the adjoint representation of SO$^{*}\left(
32\right) $ and the $527$ is the symmetric traceless tensor of SO$^{*}\left(
32\right) $. These 2080 generators form the algebra of Sp$\left( 64\right) $.

It was shown in \cite{IIB} that the B-algebra may be written in an SL$\left(
2,Z\right) $ (U-duality) basis instead of the spacetime SO$\left( 2,1\right)
=$SL$\left( 2,R\right) $ basis given above. These two bases are related to
each other by a deformation that involves the IIB string coupling constant $%
z=a+ie^{-\phi }$ (axion and dilaton moduli), and its SL$\left( 2,Z\right) $
properties $z^{\prime }=\left( az+b\right) /\left( cz+d\right) $. The
deformed B-algebra may be used to perform certain non-perturbative
computations at any value of the string coupling constant (see \cite{IIB}).
The entire superalgebra OSp$\left( 1/64\right) $ may be rewritten in the SL$%
\left( 2,Z\right) $ basis by following the prescription in \cite{IIB}.
Therefore the observations in S-theory in a IIB basis may now be interpreted
as observations in the two-time physics version of M-theory taken in a
particular gauge.

The 10D A and B bases are obviously related to each other since they both
occur in the C-branch. The map between these two corresponds to a
rearrangement of the 64 fermions that are in the spinor representation of SO$%
\left( 11,2\right) $. This map is clearly related to T-duality as discussed
in \cite{stheory}, and in the present context of two-time physics it is
interpreted as just a gauge transformation from one fixed gauge to another
fixed gauge.

The heterotic and type-I superalgebras in 10D are then obtained as in \cite
{stheory} from the IIB sector, either by setting one of the two
16-supercharges $Q_{\alpha a}$ to zero (heterotic) or by their
identification (type I). This is possible because one can work in a OSp$%
\left( 1/64\right) $ representation space labelled by the commuting
operators $P_{\bar{\mu}}^{\bar{m}}\oplus Z_{\bar{\mu}_1\cdots \bar{\mu}_5}^{%
\bar{m}}\oplus Z_{\bar{\mu}_1\bar{\mu}_2\bar{\mu}_3}$. The heterotic or
type-I sectors may be viewed as BPS-like sectors in which some of these
charges are related to each other as discussed in the second paper in \cite
{stheory}.

\section{AdS$_n\otimes S^m$ bases}

In the C-branch, starting with SO$\left( 11,2\right) $ one can come down to
the basis labelled by the subgroups SO$\left( 3,2\right) \otimes $SO$\left(
8\right) $ or to SO$\left( 6,2\right) \otimes $SO$\left( 5\right) $ , which
are the isometries of the spaces AdS$_4\times $S$^7$ and AdS$_7\times $S$^4$
respectively. The reduction is obtained by rewriting the 13D label $\tilde{M}%
=\hat{\mu}\oplus \hat{m},$ where $\hat{\mu}$, $\hat{m}$ are labels for the
vectors of SO$\left( 3,2\right) ,$SO$\left( 8\right) $ or SO$\left( 5\right)
,$SO$\left( 6,2\right) $. Also, the 64-spinor is rewritten in 32+32 form $Q_{%
\tilde{\alpha}}=\psi _{\hat{\alpha}a}^{+}\oplus \psi _{\hat{\alpha}\dot{a}%
}^{-}.$ Each 32=4$\times $8 since $\hat{\alpha}$ denotes the 4-spinor for Sp$%
\left( 4\right) \sim SO\left( 3,2\right) $ or SO$\left( 5\right) ,$ and $a,%
\dot{a}$ denote the two spinors 8$_{\pm }$ for SO$\left( 8\right) $ or SO$%
\left( 6,2\right) $. For SO$\left( 3,2\right) \otimes $SO$\left( 8\right) $
both $\psi _{\hat{\alpha}a}^{+}$ and $\psi _{\hat{\alpha}\dot{a}}^{-}$ are
real since the corresponding spinors are real. Hence, they each have 32 real
and independent components. For SO$\left( 5\right) \otimes $SO$\left(
6,2\right) $ they are in a complex basis since the 4 and the 8$_{\pm }$ are
pseudo-real. However, the pseudo-reality condition still gives 32 real and
independent components in each of the $\psi _{\alpha a}^{+},\psi _{\alpha 
\dot{a}}^{-}$. The anti-commutators in the superalgebra take the sketchy
form 
\begin{eqnarray}
\left\{ \psi _{\hat{\alpha}a}^{+},\psi _{\hat{\beta}b}^{+}\right\} &=&J_{%
\hat{\mu}\hat{\nu}}^{+}\oplus J_{+}^{\hat{m}\hat{n}}\oplus X_{\hat{\mu}}^{+%
\hat{m}\hat{n}}\oplus X_{\hat{\mu}\hat{\nu}}^{+\hat{m}_1\cdots \hat{m}_4}
\label{osp132+} \\
\left\{ \psi _{\hat{\alpha}\dot{a}}^{-},\psi _{\hat{\beta}\dot{b}%
}^{-}\right\} &=&J_{\hat{\mu}\hat{\nu}}^{-}\oplus J_{-}^{\hat{m}\hat{n}%
}\oplus X_{\hat{\mu}}^{-\hat{m}\hat{n}}\oplus X_{\hat{\mu}\hat{\nu}}^{-\hat{m%
}_1\cdots \hat{m}_4} \\
\left\{ \psi _{\hat{\alpha}a}^{+},\psi _{\hat{\beta}\dot{b}}^{-}\right\}
&=&Y^{\hat{m}}\oplus Y^{\hat{m}_1\hat{m}_2\hat{m}_3}\oplus Y_{\hat{\mu}}^{%
\hat{m}}\oplus Y_{\hat{\mu}}^{\hat{m}_1\hat{m}_2\hat{m}_3} \\
&&\oplus Y_{\hat{\mu}\hat{\nu}}^{\hat{m}}\oplus Y_{\hat{\mu}_1\hat{\mu}_2}^{%
\hat{m}_1\hat{m}_2\hat{m}_3}.  \nonumber
\end{eqnarray}
The map to the 13D operators $\tilde{J}_{2,3,6}$ can be easily established
through $\tilde{M}=\hat{\mu}\oplus \hat{m}.$ The $X_{\hat{\mu}\hat{\nu}%
}^{\pm \hat{m}_1\cdots \hat{m}_4}$ are self or anti-self dual in the
8-dimensions labelled by $m$. The operators $\psi _{\hat{\alpha}a}^{\pm }$, $%
J_2^{\pm },$ $J_{\pm }^2,$ $X_1^{\pm 2},$ $X_2^{\pm 4}$ form OSp$\left(
1/32\right) _{\pm }$ sub-supergroups, but OSp$\left( 1/32\right) _{+}$ does
not commute with OSp$\left( 1/32\right) _{-}$ since $\left\{ \psi _{\hat{%
\alpha}a}^{+},\psi _{\hat{\beta}\dot{b}}^{-}\right\} $ is not zero. However,
OSp$\left( 1/64\right) \supset $ OSp$\left( 1/32\right) _{+}\otimes $Sp$%
\left( 32\right) _{-}$. The generators $J_2^{\pm }\oplus J^{\pm 2}$ form SO$%
\left( 3,2\right) _{\pm }\otimes SO\left( 8\right) _{\pm }$ or SO$\left(
5\right) _{\pm }\otimes SO\left( 6,2\right) _{\pm }$ subgroups embedded in
each of the Sp$\left( 32\right) _{\pm }$. How is OSp$\left( 1/64\right) $
superalgebra related to the familiar AdS$\times S$ supersymmetries OSp$%
\left( 8/4\right) $ or OSp$\left( 6,2/4\right) $ ? These are not
sub-supergroups of OSp$\left( 1/64\right) $. From our analysis it can be
seen that OSp$\left( 8/4\right) $ or OSp$\left( 6,2/4\right) $ , which
includes $\psi _{\hat{\alpha}a}^{+},J_2^{+},$ $J_{+}^2$, gets enlarged by
the addition of the non-Abelian operators $X_1^{+2},X_2^{+4}$ into OSp$%
\left( 1/32\right) _{+}$ (it is not possible to set these operators to zero
naively since they are non-Abelian and they cannot be simultaneously
diagonalized in a quantum theory). In turn, OSp$\left( 1/32\right) _{+}$ is
the sector of OSp$\left( 1/64\right) $ that is a singlet under Sp$\left(
32\right) _{-}$.

We speculate that if the CFT-AdS conjecture \cite{maldacena} corresponds to
a corner of M-theory then the enlargement of the superalgebra probably does
occur on the CFT side from the point of view of the N=8 Super Yang-Mills
theory in 3D. The symmetry of this theory in perturbation theory to all
orders is OSp$\left( 8/4\right) $. As shown in \cite{super2t}, by taking the
SO$\left( 3,2\right) $ indices $\hat{\mu}=$ $\mu ,\pm ^{\prime }$ and $\psi
_{\hat{\alpha}a}^{+}\sim Q_{\alpha a}^{1/2}\oplus S_{\alpha a}^{-1/2}$, the
conformal supersymmetry of the AdS$_4\times S^7$ background can be rewritten
in the compact form of Eq.(\ref{osp132+}), excluding the $X_1^{+2}$ and $%
X_2^{+4}$ $\,$generators. How can one see the enlargement to OSp$\left(
1/32\right) _{+}$? One begins with non-perturbative field configurations
that turn on the central extensions $\left\{ Q_{\alpha a}^{1/2},Q_{\beta
b}^{1/2}\right\} \sim X_{\hat{\mu}=+^{\prime }}^{+\hat{m}\hat{n}}$; then the
conformal symmetry $K_\mu $ requires all $X_{\hat{\mu}=-^{\prime }}^{+\hat{m}%
\hat{n}}$, $X_{\hat{\mu}=\mu }^{+\hat{m}\hat{n}},$ and their non-Abelian
nature generates $X_{\hat{\mu}\hat{\nu}}^{+\hat{m}_1\cdots \hat{m}_4}$, thus
completing the OSp$\left( 1/32\right) _{+}$ superalgebra of Eq.(\ref{osp132+}%
). This argument shows that the inclusion of non-perturbative physics in the
AdS$_4\times $S$^7$ background could be described by the Sp$\left( 32\right)
_{-}$-singlet sector of OSp$\left( 1/64\right) $. The singlet sector can
arise as a result of a contraction that would be related to the limits \cite
{maldacena} one must take to establish the AdS-CFT correspondance.

Similarly, in the C-branch, starting with SO$\left( 11,2\right) $ we can
come down to a 12D$^{\prime }$ basis by separating the 13th spacelike
dimension $\tilde{M}=M\oplus 1^{\prime }$. This gives 64=32$_L+32_R$ where 32%
$_{L,R}$ are the two SO$^{\prime }\left( 10,2\right) $ spinors $\psi ^L$, $%
\psi ^R$ of {\it opposite} chirality (contrast 12D$^{\prime }$ to the 12D of
the A-branch which gave {\it same} chirality). Next we separate 12D$^{\prime
}$=6D+6D so that SO$^{\prime }\left( 10,2\right) \rightarrow SO\left(
6,2\right) \otimes SO\left( 6\right) $ which is the isometry group of the AdS%
$_5\times S^5$ space. Consider the 32$_L$ real components $\psi _{\alpha
a}^L $ written in the $\left( 4,4\right) $ complex spinor basis with $\alpha 
$ and $a$ denoting the complex 4 spinors of SU$\left( 2,2\right) =SO\left(
4,2\right) $ and SU$\left( 4\right) =SO\left( 6\right) $ respectively. Since
the basis is complex we need to consider the hermitian conjugates $\left(
\psi _{\alpha a}^{L,R}\right) ^{\dagger }\equiv \bar{\psi}_{\dot{\alpha}\dot{%
a}}^{L,R}$ where $\dot{\alpha}$ and $\dot{a}$ denote the complex \={4}
spinors. Then the 12D$^{\prime }$ basis of OSp$\left( 1/64\right) $ is
reduced to the 6D+6D basis and it takes the following sketchy form 
\begin{eqnarray}
\left\{ \psi _{\alpha a}^L,\bar{\psi}_{\dot{\beta}\dot{b}}^L\right\} &\sim
&J_2^L\oplus J_L^2\oplus J_L\oplus X_2^{L2},\quad \\
\left\{ \psi _{\alpha a}^L,\psi _{\beta b}^L\right\} &\sim &Z_1^{L1}\oplus
Z_3^{L3},\,\,\quad and\,\,\,\,\,h.c. \\
\left\{ \psi _{\alpha a}^R,\bar{\psi}_{\dot{\beta}\dot{b}}^R\right\} &\sim
&J_2^R\oplus J_R^2\oplus J_R\oplus X_2^{R2},\quad \\
\left\{ \psi _{\alpha a}^R,\psi _{\beta b}^R\right\} &\sim &Z_1^{R1}\oplus
Z_3^{R3},\,\,\quad \,and\,\,\,\,h.c., \\
\left\{ \psi _{\alpha a}^L,\bar{\psi}_{\dot{\beta}\dot{b}}^R\right\} &\sim
&Y_2\oplus Y^2\oplus Y\oplus Y_2^2,\quad and\,\,\,h.c. \\
\left\{ \psi _{\alpha a}^L,\psi _{\beta b}^R\right\} &\sim &W_1^1\oplus
W_1^3\oplus W_3^1\oplus W_3^3,\,\,\quad and\,\,\,\,h.c.
\end{eqnarray}
where the generators $J_p^q,X_p^q,Y_p^q,Z_p^q,W_p^q$ are labelled with
numbers in subscripts or superscripts that are SO$\left( 4,2\right) $
p-forms or SO$\left( 6\right) $ q-forms, and $h.c.$ stand for hermitian
conjugate relations. The map to the 13D operators $\tilde{J}_{2,3,6}$ can be
easily established through the reduction $\tilde{M}=\tilde{\mu}\oplus \tilde{%
m}\oplus 1^{\prime }$ where $\tilde{\mu},\tilde{m}$ are labels for the
vectors of SO$\left( 4,2\right) ,$SO$\left( 6\right) $.

$J_2^L\oplus J_L^2$ generate SU$\left( 2,2\right) \otimes SU\left( 4\right) $
which is the isometry of AdS$_5\times S^5,$ i.e. SO$\left( 4,2\right)
\otimes SO\left( 6\right) \subset SO^{\prime }\left( 10,2\right) $. The
supersymmetry algebra in this background is $SU(2,2/4)$. What is the
relation of $OSp(1/64)$ and $SU(2,2/4)$? As before, this is not a
sub-supergroup. Again, upon the inclusion of the charges $J_L\oplus
X_2^{L2}\oplus Z_1^{L1}\oplus Z_3^{L3}$ it can be seen that $SU(2,2/4)$ is
enlarged into $OSp(1/32)_L$ as in the first two lines of the equations above
(see also \cite{craps} \cite{ferrara}). Then we see that SU$\left(
1/64\right) \supset OSp(1/32)_L\otimes Sp\left( 32\right) _R$ and the sector
that is singlet under $Sp\left( 32\right) _R$ is described by $OSp(1/32)_L$.

If one starts with the superconformal N=4 super Yang-Mills theory in 4D, the
perturbative symmetry is SU$\left( 2,2/4\right) $. By including central
extensions that correspond to non-perturbative backgrounds such as monopoles
and dyons one turns on the central charge $Z_{+^{\prime }}^{L\tilde{m}}$
which is part of $Z_1^{L1}$. Conformal symmetry requires the full $Z_1^{L1}$
and its hermitian conjugate $\left( Z_1^{L1}\right) ^{\dagger }$. Their
commutators generate all the other remaining charges to complete the OSp$%
(1/32)_L$ superalgebra. Thus the inclusion of non-perturbative physics in
the AdS$_5\times $S$^5$ background could be described by the Sp$\left(
32\right) _R$-singlet sector of OSp$\left( 1/64\right) $. The singlet sector
can arise as a result of a contraction that may be related to the limits 
\cite{maldacena} one must take to establish the AdS-CFT correspondance.

\section{Toy M-Model, cousins of kappa symmetry}

Our ideas can be dynamically illustrated with a toy M-model on the worldline
with a new set of gauge symmetries. The main point is that in this model the
various dual bases described above emerge naturally by making appropriate
gauge choices within the same theory. Conversely, one can transform one
basis to another dual basis by making gauge transformations, thus imitating
duality in M-theory. Here we only outline the general structure of such a
model, leaving the details to a future paper \cite{future}. Similar
structures could be constructed for any group $G$ and a choice of subgroup $%
H $ as outlined at the end of \cite{super2t}.

Consider the supergroup $G=$OSp$\left( 1/64\right) $ with the subgroup $H=$SO%
$\left( 11,2\right) $. Let $X_i^{\tilde{M}}=(X^{\tilde{M}},P^{\tilde{M}})$
represent position / momentum SO$\left( 11,2\right) $ vectors. The $0$-brane
vectors $X_i^{\tilde{M}}\left( \tau \right) $ form an Sp$\left( 2\right) $
doublet. Sp$\left( 2\right) $ is gauged by including the gauge potentials $%
A^{ij}\left( \tau \right) $. This gauge symmetry introduces first class
constraints whose solution requires two-timelike dimensions as explained in 
\cite{lifting}. In addition, consider the 65$\times 65$ matrix which is a
group element $g\left( \tau \right) \in $OSp$\left( 1/64\right) $. It is a
singlet under Sp$\left( 2\right) $. The subgroup $H=$SO$\left( 11,2\right) $
acting simultaneously on the {\it left side} of $g$ and on the vectors $X_i^{%
\tilde{M}}$ is gauged. The gauge potential is $\Omega ^{\tilde{M}\tilde{N}%
}\left( \tau \right) $. With this information we can write covariant
derivatives as in \cite{super2t} 
\begin{eqnarray}
D_\tau X_i^{\tilde{M}} &=&\partial _\tau X_i^{\tilde{M}}-\varepsilon
_{ik}A^{kj}X_i^{\tilde{M}}-\Omega ^{\tilde{M}\tilde{N}}X_{i\tilde{N}}\,, \\
D_\tau g &=&\partial _\tau g-\frac 14\Omega ^{\tilde{M}\tilde{N}}\left(
\Gamma _{\tilde{M}\tilde{N}}g\right) \,.
\end{eqnarray}
Consider the part of the Cartan connection $\left( D_\tau g\right) g^{-1}$
restricted to the subgroup $H$%
\begin{equation}
\left( D_\tau gg^{-1}\right) _H=\frac 1{32}Str\left( \Gamma _{\tilde{M}%
\tilde{N}}\,D_\tau gg^{-1}\right) =\frac 1{32}Str\left( \Gamma _{\tilde{M}%
\tilde{N}}\,\partial _\tau gg^{-1}\right) -\Omega _{\tilde{M}\tilde{N}}
\end{equation}
A Lagrangian that is invariant under the gauge symmetry Sp$\left( 2\right)
\times SO\left( 11,2\right) $ is given by 
\begin{equation}
\pounds =\frac 12\,\varepsilon ^{ij}D_\tau X_i^{\tilde{M}}X_j^{\tilde{N}%
}\eta _{\tilde{M}\tilde{N}}+\frac 12\left[ \left( D_\tau gg^{-1}\right)
_H\right] ^2+\frac 12\left( \varepsilon ^{ij}X_i^{\tilde{M}}X_j^{\tilde{N}%
}\right) ^2.
\end{equation}
This action is invariant under the {\it global} symmetry OSp$\left(
1/64\right) $ that acts {\it linearly} on the {\it right side} of $g\left(
\tau \right) $. On the left side of $g\left( \tau \right) $ the evident
symmetry is smaller $H$=SO$\left( 11,2\right) $ because of the presence of $%
\Omega $, but as we will see there is a much bigger local symmetry that is
realized non-linearly. Thus the model is realized in the C-branch OSp$\left(
1/64\right) $. As we saw above, the C-branch contains all the interesting
spacetime interpretations at its intersections with the A and B branches and
with the OSp$\left( 1/32\right) $ branch.

The group element $g\left( \tau \right) $ can be written in the form $g=ht$
where $h\in H$ and $t\in G/H$. We can use the $H=SO\left( 11,2\right) $
gauge symmetry to choose a unitary gauge by eating away $h\left( \tau
\right) $. Using the equations of motion (or doing the path integral) one
may solve for $\Omega _{\tilde{M}\tilde{N}}=\tilde{\Omega}_{\tilde{M}\tilde{N%
}}+L_{\tilde{M}\tilde{N}}$ , with $\tilde{\Omega}_{\tilde{M}\tilde{N}}=\frac
1{32}Str\left( \Gamma _{\tilde{M}\tilde{N}}\partial _\tau tt^{-1}\right) $
and substitute back into the action to find the version of the action given
in \cite{super2t} 
\begin{equation}
\pounds =\,\partial _\tau X_1\cdot X_2-A^{ij}X_i\cdot X_j-\frac
1{32}Str\left( \Gamma _{\tilde{M}\tilde{N}}\partial _\tau tt^{-1}\right) L^{%
\tilde{M}\tilde{N}}.
\end{equation}
where $L_{\tilde{M}\tilde{N}}=\varepsilon ^{ij}X_i^{\tilde{M}}X_j^{\tilde{N}%
}=X^{\tilde{M}}P^{\tilde{N}}-X^{\tilde{N}}P^{\tilde{M}}$ is the orbital
angular momentum in 13D. A total derivative has been dropped in the first
term of $\pounds $. With this we see that the canonical formalism gives $%
X_1^M=X^M$ as the position and $X_2^{\tilde{M}}=P^{\tilde{M}}$ as the
momentum of the 13D zero-brane.

We parametrize $t\left( a\left( \tau \right) ,\Theta \left( \tau \right)
\right) $ it terms of 64 fermionic $\Theta _{\tilde{\alpha}}\left( \tau
\right) $ and 286+1716 bosonic $a_{\tilde{M}_1\tilde{M}_2\tilde{M}_3}\left(
\tau \right) ,$ $a_{\tilde{M}_1\cdots \tilde{M}_6}\left( \tau \right) $
degrees of freedom that correspond to 13D p-forms with p=3,6. As described
in \cite{super2t}, the global symmetry $g\subset $OSp$\left( 1/64\right) $
acts on the {\it right side} of $t\left( \tau \right) $ and it must be
compensated by a field dependent, gauge restoring transformation $%
h^{-1}\left( a,\Theta ;g\right) $ on the left side 
\begin{equation}
t\left( a,\Theta \right) \rightarrow t\left( a^{\prime },\Theta ^{\prime
}\right) =h^{-1}t\left( a,\Theta \right) g\,\,.
\end{equation}
$h\left( \tau \right) $ acts as a field dependent SO$\left( 11,2\right) $
gauge transformation on $\tilde{\Omega}^{\tilde{M}\tilde{N}}\left( a,\Theta
\right) $ and $X_i^{\tilde{M}}$. The action remains invariant because it was
built as a gauge invariant under SO$\left( 11,2\right) $ and globally
invariant under OSp$\left( 1/64\right) $.

The action in this gauge involves the following fields on the worldline $%
X_i^{\tilde{M}}\left( \tau \right) ,$ $A^{ij}\left( \tau \right) ,$ $%
a_3\left( \tau \right) ,$ $a_6\left( \tau \right) ,$ $\Theta \left( \tau
\right) $. The $X_i^{\tilde{M}}\left( \tau \right) ,\Theta ^{\tilde{\alpha}%
}\left( \tau \right) $ may be interpreted as the 13D superspace of two-time
physics as in \cite{super2t}, however the bigger dimensions in the present
case, as opposed to the 3,4,6D cases of \cite{super2t}, require the extra
bosonic degrees of freedom $a_3,a_6$. From the superalgebra point of view
these are closely associated with the $p$-brane charges $\tilde{J}_3,\tilde{J%
}_6$ of the C-branch, hence the $a_p$ may be thought of as the $\tau $%
-component of $\left( p+1\right) $-form gauge potentials $A_{p+1}$ for $%
p=3,6 $.

There is also additional local symmetry beyond SO$\left( 11,2\right) $ that
originates with transformations on the left side of $t\left( a,\Theta
\right) $ with fermionic as well as bosonic parameters $b_3(\tau ),b_6\left(
\tau \right) ,\kappa ^{\tilde{\alpha}}\left( \tau \right) $ that form a
generalization of kappa supersymmetry. The transformation is 
\begin{equation}
t\left( a,\Theta \right) \rightarrow t\left( a^{^{\prime \prime }},\Theta
^{^{\prime \prime }}\right) =h^{-1}t\left( b,\kappa \right) \,t\left(
a,\Theta \right) ,
\end{equation}
where again $h\left( b,\kappa ;a,\Theta \right) $ is the induced local
Lorentz transformation. For infinitesimal $b,\kappa $ we have $%
t=1_{65}+b_3\Gamma ^3+b_6\Gamma ^6+\kappa _{\tilde{\alpha}}F^{\tilde{\alpha}%
} $ where $\Gamma ^{3,6}$ and $F^{\tilde{\alpha}}$ are 65$\times 65$
matrices that provide representations of the generators of $G/H$. Then we
find that the Lagrangian is invariant under these additional gauge
symmetries provided the parameters $\kappa ,b_3,b_6$ are constrained by the
following relations (where $\sim X_i\cdot X_j$ means proportional up to
field dependent functions with appropriate indices) 
\begin{equation}
L_{\tilde{M}\tilde{N}}\left( \Gamma ^{\tilde{M}\tilde{N}}\kappa \right) _{%
\tilde{\alpha}}\sim L_{[\tilde{M}_1}^{\,\,\tilde{N}}b_{\tilde{M}_2\tilde{M}%
_3]\tilde{N}}\sim L_{[\tilde{M}_1}^{\,\,\tilde{N}}b_{\tilde{M}_2\cdots 
\tilde{M}_6]\tilde{N}}\sim X_i\cdot X_j.
\end{equation}
The left side looks like Lorentz transformations on $\kappa ,b_3,b_6$ with a
Lorentz parameter $L_{\tilde{M}\tilde{N}}$ . Then the part proportional to $%
X_i\cdot X_j$ is cancelled by choosing $\delta A^{ij}$ so that \pounds\ is
invariant (see \cite{super2t} for an explicit example). These gauge
symmetries are more than sufficient to remove the ghosts in $\Theta ,a_3,a_6$
associated with the extra timelike dimension in 13D. We will call these
gauge symmetries ``extended kappa symmetries''.

We may now make various gauge choices for Sp$\left( 2\right) ,$ SO$\left(
11,2\right) $, and extended kappa symmetries, that are manifestly covariant
under various spacetime interpretations (one-time) as discussed in the
purely algebraic discussion in the first part of this paper. To do this we
reorganize the degrees of freedom $X_i^{\tilde{M}}\left( \tau \right) ,$ $%
a_3\left( \tau \right) ,$ $a_6\left( \tau \right) ,$ $\Theta _{\tilde{\alpha}%
}\left( \tau \right) $ according to the representations of the various
subgroups of SO$\left( 11,2\right) $ as we did for the generators of OSp$%
\left( 1/64\right) $. Some of the pieces of $X_i,a_3,a_6,\Theta $ are set to
zero or constants by gauge fixing. Then part of OSp$\left( 1/64\right) $ is
realized linearly on the remaining degrees of freedom and the remainder is
realized non-linearly. For such examples in simpler cases 
without supersymmetry see \cite{lifting}
and with supersymmetry see \cite{super2t}. This discussion
makes it evident that we can choose gauges that would realize the 11D
M-algebra, the 10D type IIA, type IIB, heterotic, type-I supersymmetries, or
the superalgebras describing the $AdS_n\times S^m$ backgrounds. These gauge
choices provide different looking toy models that have the supersymmetries
of the various corners of the moduli space of M-theory \cite{future}. Since
they are all derived from the same unified 13D theory by gauge choices, they
can be transformed into each other by gauge transformations that correspond
to dualities in M-theory.

The toy M-model given here has a much richer set of possible gauge choices
along the lines of \cite{lifting} that tie together systems such as free
particles, Hydrogen atom, harmonic oscillators, curved spaces, and even
non-relativistic systems with more general potentials, etc. This suggests
that M-theory may have similar properties. It appears that M-theory could be
lifted to two-time physics with a global symmetry OSp$\left( 1/64\right) $.
We are hopeful that M-theory will be better understood by studying it
covariantly in 13D in the formalism of two-time physics in the presence of
various gauge degrees of freedom and new gauge symmetries of the type
described here.

\end{document}